# Sensitivity of magnetic properties to chemical pressure in lanthanide garnets $Ln_3A_2X_3O_{12}$, $Ln$ = Gd, Tb, Dy, Ho, $A$ = Ga, Sc, In, Te, $X$ = Ga, Al, Li

P Mukherjee[1*], A C Sackville Hamilton[2], H F J Glass[1], S E Dutton[1*]

[1] Cavendish Laboratory, University of Cambridge, JJ Thomson Avenue, Cambridge CB3 0HE, United Kingdom

[2] London Centre for Nanotechnology, University College London, London WC1H 0AH, United Kingdom

*Email: pm545@cam.ac.uk, sed33@cam.ac.uk

**Abstract**

A systematic study of the structural and magnetic properties of three-dimensionally frustrated lanthanide garnets $Ln_3A_2X_3O_{12}$, $Ln$ = Gd, Tb, Dy, Ho, $A$ = Ga, Sc, In, Te, $X$ = Ga, Al, Li is presented. Garnets with $Ln$ = Gd show magnetic behaviour consistent with isotropic $Gd^{3+}$ spins; no magnetic ordering is observed for $T \geq 0.4$ K. Magnetic ordering features are seen for garnets with $Ln$ = Tb, Dy, Ho in the temperature range $0.4 < T < 2.5$ K, however the nature of the magnetic ordering varies for the different $Ln$ as well as for different combinations of $A$ and $X$. The changes in magnetic behaviour can be explained by tuning of the magnetic interactions and changes in the single-ion anisotropy. The change in magnetic entropy is evaluated from isothermal magnetisation measurements to characterise the magnetocaloric effect in these materials. Among the Gd garnets, the maximum change in magnetic entropy per mole (15.45 J K$^{-1}$ mol$_{Gd}^{-1}$) is observed for $Gd_3Sc_2Ga_3O_{12}$ at 2 K, in a field of 9 T. The performance of $Dy_3Ga_5O_{12}$ as a magnetocaloric material surpasses the other garnets with $Ln$ = Tb, Dy, Ho.

# 1 Introduction

Magnetic lattices where the underlying lattice geometry prevents all the pairwise magnetic interactions from being satisfied simultaneously are said to be geometrically frustrated. In such systems, a delicate balance among competing interactions determines the magnetic ground state[1,2]. Geometrically frustrated lanthanide oxides are known to exhibit exotic magnetic behaviour such as spin ice or spin liquid ground states[3–8]. The magnetic properties of such materials can vary radically even with small changes in the structure. One way of systematically studying such variations is by exploring the effect of chemical pressure, that is, changing the size of the nonmagnetic cation in the lattice. The effect of doing so is twofold: a) it alters the lattice parameter and hence the distances between the magnetic $Ln^{3+}$, which changes the dipolar and exchange interactions b) it causes subtle changes to the $Ln$-O environment, which affects the crystal electric field (CEF) and hence the single-ion anisotropy of the magnetic $Ln^{3+}$ as well as the superexchange interactions. Such studies have been extensively carried out for the highly frustrated lanthanide pyrochlores, $Ln_2B_2O_7$ ($B$ = non-magnetic cation), where the dominant interactions are the nearest-neighbour exchange, the dipolar interaction and CEF effects. While the pyrochlores with $Ln$ = Dy, Ho remain in the spin ice state for different $B$ cations, varying the chemical pressure radically changes the magnetic ground state for $Ln$ = Tb, Yb, Gd[9–15].

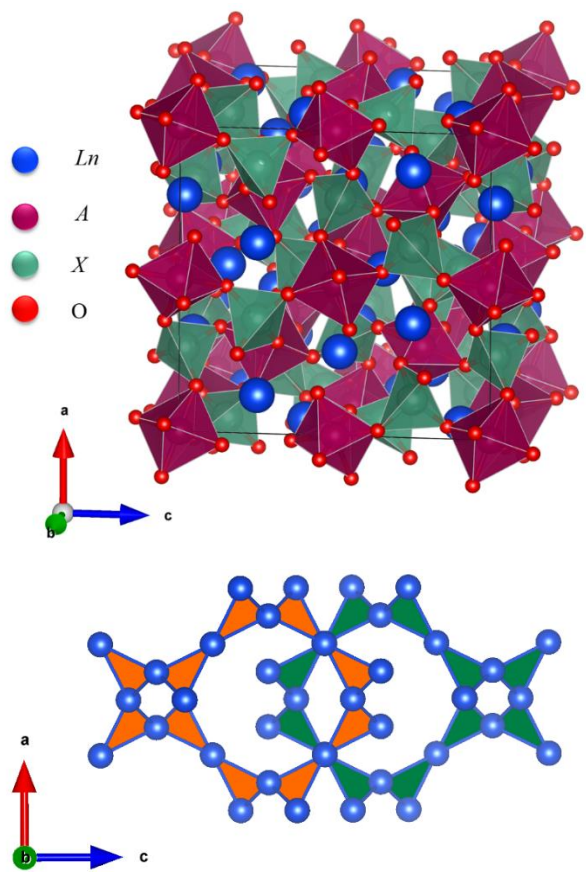

Figure 1 - a) Crystal structure of lanthanide garnets $Ln_3A_2X_3O_{12}$ b) Connectivity of magnetic $Ln^{3+}$ ions: The $Ln^{3+}$ lie at the vertices of corner-sharing equilateral triangles forming two interpenetrating ten-membered rings. This results in a highly frustrated three-dimensional network.

In this paper, we explore the effect of chemical pressure on another well-known three-dimensional frustrated system, the lanthanide garnets, having the general formula $Ln_3A_2X_3O_{12}$. $A$ and $X$ are both restricted to non-magnetic cations as inclusion of magnetic cations on the $A$ and $X$ sites gives rise to additional magnetic phenomena[16–21]. The garnets crystallise in a cubic structure and contain three distinct cation sites based on the coordination with oxygen: dodecahedral occupied by $Ln$, octahedral occupied by $A$ and tetrahedral occupied by $X$, Figure 1a. The magnetic $Ln^{3+}$ ions are located at the vertices of corner-sharing triangles which form two interpenetrating networks of bifurcated ten membered rings, giving rise to geometrical frustration, Figure 1b[22,23]. As there are two non-magnetic cation sites in the lanthanide garnets, there is a lot of potential for exploring the magnetic phase diagram in these garnets by varying the size of the ions and hence the chemical pressure on either or both $A$ and $X$ sites. Changing the chemical pressure allows tuning of the magnetic interactions as well as changes to the single-ion anisotropy, offering an opportunity to study the physics of frustration in lanthanide garnets. Any changes in magnetic properties would also impact the magnetocaloric effect (MCE) in these materials. This is significant because gadolinium gallium garnet (GGG), $Gd_3Ga_5O_{12}$, and dysprosium gallium garnet (DGG), $Dy_3Ga_5O_{12}$, are standard magnetocaloric materials (MCMs) for solid state magnetic refrigeration in the liquid helium temperature regime[24,25]. Therefore, it may be possible to optimise the MCE by varying the cations on the $A$ and $X$ sites. This has been reported for Al substituted GGG[26], however the impact on the MCE for the other $Ln^{3+}$ as well as for other combinations of $A$ and $X$ have not been explored.

Before we embark on our study, we review the current state of knowledge regarding the magnetic properties of the lanthanide garnets with nonmagnetic cations on the $A$ and $X$ sites. We specifically focus on the lanthanide ions considered in this study, $Ln$ = Gd, Tb, Dy, Ho.

$Gd_3A_2X_3O_{12}$:

Most of the experimental work on lanthanide garnets has focused on GGG, where both the octahedral and tetrahedral sites are occupied by $Ga^{3+}$. GGG is a canonical spin liquid candidate with no long-range ordering down to 0.025 K and a glassy transition at $T_g \sim 0.14$ K below the spin liquid state[8,27–32]. Recent experiments have pointed to the existence of a hidden multipolar order on the ten-membered loops in the spin liquid state[32], as well as existence of dispersionless spin waves on the ten-membered loops in high magnetic fields[30]. A study comparing the magnetic properties of GGG, $Gd_3Al_5O_{12}$ (with both octahedral and tetrahedral sites occupied by $Al^{3+}$) and $Gd_3Te_2Li_3O_{12}$ (with octahedral sites occupied by $Te^{6+}$ and tetrahedral sites occupied by $Li^+$) reports a sharp transition at 0.243 K for $Gd_3Te_2Li_3O_{12}$ and a broader transition at 0.175 K for $Gd_3Al_5O_{12}$, in contrast to no ordering for GGG. The differences in magnetic behaviour are attributed to a subtle variation in the ratio of dipolar and nearest-neighbour exchange interactions, with the gradual increase in dipolar interaction from $Gd_3Al_5O_{12}$ to $Gd_3Te_2Li_3O_{12}$ relieving the frustration[33].

$Ln_3A_2X_3O_{12}$, $Ln$ = Tb, Dy, Ho:

$Ln_3Ga_5O_{12}$ ($Ln$ = Tb, Dy, Ho) undergo transitions at much lower temperatures as compared to the corresponding $Ln_3Al_5O_{12}$, and thus, much like their Gd counterparts, are more frustrated[33,34]. $Dy_3Al_5O_{12}$ has a strong Ising anisotropy with a sharp antiferromagnetic ordering transition at $T_N = 2.49$ K; neutron diffraction experiments report ordering in the so called six sublattice antiferromagnetic type A (AFA) structure below $T_N$[35–37]. The same

AFA ordering occurs in $Tb_3Al_5O_{12}$ and $Ho_3Al_5O_{12}$ at $T_N = 1.35$ K and $T_N = 0.95$ K[38–40]. The situation for the $Ln_3Ga_5O_{12}$ ($Ln$ = Tb, Dy, Ho) is more complex. Bulk magnetic measurements and neutron diffraction show that $Tb_3Ga_5O_{12}$ and $Ho_3Ga_5O_{12}$ order antiferromagnetically below $T_N = 0.25$ K and $T_N = 0.19$ K in the AFA structure; however the ordering mechanism involves hyperfine interactions in addition to the dipolar and exchange interactions[41–44]. Crystal field studies suggest a quasi-planar anisotropy for $Dy_3Ga_5O_{12}$[45], in contrast to the strongly Ising nature of $Dy_3Al_5O_{12}$. Heat capacity measurements show a broad short-range ordering at 0.6-0.7 K and a sharp transition at $T_N = 0.373$ K, neutron diffraction measurements at 0.07 K report antiferromagnetic ordering below $T_N$[46,47]. A later neutron scattering experiment on $Ho_3Ga_5O_{12}$ reports the onset of short-range order below 0.6 K and coexistence of long and short-range order below 0.3 K down to 0.05 K[48]. This behaviour is similar to that reported in the bulk measurements for $Dy_3Ga_5O_{12}$[47], however in both cases, the exact nature of the short-range order is yet to be elucidated. The increased transition temperatures in $Ln_3Al_5O_{12}$ ($Ln$ = Tb, Dy, Ho) have been attributed to increased dipolar interactions and stronger Ising anisotropy[38]. A previous study on the lanthanide tellurate lithium garnets, $Ln_3Te_2Li_3O_{12}$, with aliovalent $A$ and $X$, reports a transition at 2 K for $Ln$ = Dy; no transition is reported for $Ln$ = Tb and Ho at $T \geq 2$ K[49].

In this work, we report on the synthesis, characterisation and bulk magnetic properties of polycrystalline samples of $Ln_3Ga_5O_{12}$, $Ln_3Sc_2Ga_3O_{12}$, $Ln_3Sc_2Al_3O_{12}$, $Ln_3In_2Ga_3O_{12}$ and $Ln_3Te_2Li_3O_{12}$ for $Ln$ = Gd, Tb, Dy, Ho. Magnetic susceptibility and isothermal magnetisation measurements have been carried out to study the magnetic behaviour for $T \geq 2$ K, while zero field heat capacity measurements have been carried out to investigate the existence of magnetic ordering transitions for $T \geq 0.4$ K. The change in magnetic entropy has been evaluated to characterise the MCE. The magnetic properties and the degree of magnetic frustration are discussed in relation to the reported literature for $Ln_3Ga_5O_{12}$. Varying the chemical pressure is seen to have a dramatic impact on the magnetic behaviour of the lanthanide garnets.

## 2 Experimental Section

Powder samples of $Ln_3A_2X_3O_{12}$ ($Ln$ = Gd, Tb, Dy, Ho; $A$ = Ga, Sc, In, Te; $X$ = Ga, Al, Li) were prepared using a solid-state synthesis. Samples of $Ln_3Sc_2Ga_3O_{12}$ and $Ln_3In_2Ga_3O_{12}$ were prepared by mixing stoichiometric amounts of $Ln_2O_3$ ($Ln$ = Gd, Dy, Ho) or $Tb_4O_7$, $Ga_2O_3$ and $Sc_2O_3$ or $In_2O_3$. To ensure the correct stoichiometry, $Gd_2O_3$ and $Ga_2O_3$ were pre-dried at 800 °C and 500 °C respectively. Pellets were heated at increasingly higher temperatures between 1200 – 1350 °C for 48-72 hours with intermittent re-grindings. For $Ln_3Sc_2Al_3O_{12}$ an alternative synthesis route was followed to prevent the formation of a $LnAlO_3$ perovskite impurity phase. The starting materials $Ln(NO_3)_3$ ($Ln$ = Gd, Tb, Dy, Ho) and $Al(NO_3)_3$ were dried overnight at 80 °C and 60 °C respectively to remove any excess water of crystallisation. Stoichiometric amounts of $Ln(NO_3)_3$ ($Ln$ = Gd, Tb, Dy, Ho), $Al(NO_3)_3$ and $Sc_2O_3$ were mixed intimately. Following a pre-reaction at 1000 °C, pellets were heated at increasingly higher temperatures between 1200 – 1400 °C between 48-72 hours with intermittent re-grindings. Samples of $Ln_3Te_2Li_3O_{12}$ and $Ln_3Ga_5O_{12}$ were prepared as described elsewhere[26,33].

Formation of phase pure products was confirmed using powder X-Ray diffraction (PXRD). Short scans were initially collected over the angular range $5° \leq 2\theta \leq 60°$ using a Panalytical Empyrean X-Ray diffractometer (Cu K$\alpha$ radiation, $\lambda = 1.540$ Å) to track the progress of the reaction. For quantitative structural analysis, longer scans for 2 hours over a wider angular range $5° \leq 2\theta \leq 90°$ were collected on the same instrument. For garnets with $Ln$ = Ho (except Ho$_3$Ga$_5$O$_{12}$), room temperature (RT) powder neutron diffraction (PND) experiments were carried out on the D2B diffractometer, ILL ($\lambda = 1.595$ Å) at 300 K. Rietveld refinement was carried out using the Fullprof suite of programs[50]. Backgrounds were fitted using linear interpolation and the peak shape was modelled using a pseudo-Voigt function.

Magnetic susceptibility measurements were carried out on a Quantum Design Magnetic Properties Measurement System (MPMS) with a Superconducting Quantum Interference Device (SQUID) magnetometer. The zero-field cooled (ZFC) susceptibility was measured in a field of 100 Oe in the temperature range 2-300 K. The magnetisation, $M$, varies linearly with the magnetic field, $H$, in a field of 100 Oe and so the approximation for molar susceptibility, $\chi(T) \sim M/H$ is valid. Isothermal magnetisation $M(H)$ measurements were carried out at specific temperatures in the field range 0 – 9 T using the ACMS (AC measurement system) option on a Quantum Design Physical Properties Measurement System (PPMS).

Heat capacity measurements were performed in zero field from 0.4 – 10 K using the He3 option on a Quantum Design PPMS. Equal amounts of the sample and silver powder (99.99% Alfa Aesar) were mixed and pressed into a pellet which was then used for measurement. The contribution from silver was subtracted using values from the literature[51] in order to obtain the sample heat capacity. The magnetic heat capacity, $C_{mag}$, was obtained by subtracting the lattice contribution using a Debye model[52], with Debye temperatures ranging between 285 – 420 K.

## 3 Results

### 3.1 Crystal Structure

PXRD indicated the formation of lanthanide garnets $Ln_3$Ga$_5$O$_{12}$, $Ln_3$Sc$_2$Ga$_3$O$_{12}$, $Ln_3$Sc$_2$Al$_3$O$_{12}$, $Ln_3$In$_2$Ga$_3$O$_{12}$ and $Ln_3$Te$_2$Li$_3$O$_{12}$ ($Ln$ = Gd, Tb, Dy, Ho). Attempts to substitute Sc$^{3+}$ and In$^{3+}$ on the $X$ site resulted in the formation of $Ln$ScO$_3$ and $Ln$InO$_3$ impurities. We postulate that Sc$^{3+}$ and In$^{3+}$ are not stable in the tetrahedral coordination. Attempts to synthesise $Ln_3$ScGa$_4$O$_{12}$ yielded a mixed phase of $Ln_3$Sc$_2$Ga$_3$O$_{12}$ and $Ln_3$Ga$_5$O$_{12}$. Synthesis of $Ln_3$InGa$_4$O$_{12}$ was not attempted. Attempts to synthesise $Ln_3$In$_2$Al$_3$O$_{12}$ by solid state as well as sol-gel methods resulted in the formation of $Ln$AlO$_3$, $Ln$InO$_3$ and In$_2$O$_3$ impurities and it was concluded that this synthesis was not possible due to the large difference in the size of In$^{3+}$ and Al$^{3+}$ ions[53].

RT structural refinements for garnets with $Ln$ = Gd, Tb, Dy were carried out using PXRD only. Combined RT PXRD + PND structural refinements were carried out for garnets with $Ln$ = Ho, except Ho$_3$Ga$_5$O$_{12}$. Figure 2 shows the combined refinement for Ho$_3$Sc$_2$Al$_3$O$_{12}$. All the garnets crystallise in the same cubic structure with space group $Ia\bar{3}d$. The magnetic $Ln^{3+}$ ($Ln$

= Gd, Tb, Dy, Ho) occupy the dodecahedral 24c (0, 0.25, 0.125) site. $Ga^{3+}$, $Sc^{3+}$, $In^{3+}$ or $Te^{6+}$ occupy the octahedral 16a (0, 0, 0) site while the tetrahedral 24d (0, 0.25, 0.375) site is occupied by $Ga^{3+}$, $Al^{3+}$ or $Li^+$ in the respective garnets. $O^{2-}$ occupies the 96h (x, y, z) site. Structural parameters are summarised in Table 1. Rietveld analysis shows that the lattice volume varies linearly with the ionic radius of the $Ln^{3+}$ ion for a given combination of A and X ions (Figure 3a). A similar relationship was expected between the lattice volume and the weighted ionic radii of the A and X ions for each $Ln^{3+}$, $r_{av} = \frac{2r_A + 3r_X}{5}$, however as is seen in Figure 3b, this is not the case for all combinations of A and X. While the $Ln_3A_2Ga_3O_{12}$ family (A = Ga, Sc, In) follows a linear trend for all Ln, the $Ln_3Sc_2Al_3O_{12}$ and $Ln_3Te_2Li_3O_{12}$ family deviate from the straight line implying that there are other effects to be considered such as the difference in sizes or charges of the A and X sites.

Selected bond lengths are given in Table 2. The changes in Ln-Ln bond lengths follow the same trend as the changes in the lattice parameters, as expected. For a particular Ln, there are subtle changes in the Ln-O bond lengths which may impact the local CEF and hence the single ion anisotropy of the magnetic $Ln^{3+}$.

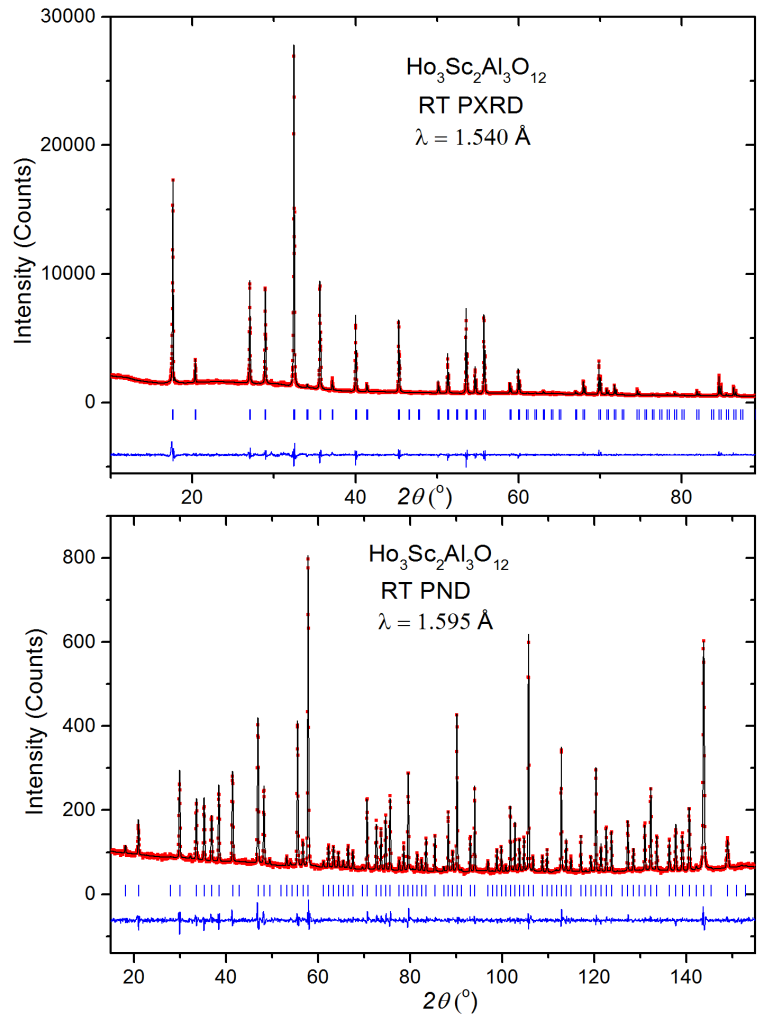

Figure 2 – RT PXRD + PND pattern for $Ho_3Sc_2Al_3O_{12}$: Experimental data (red dots), Modelled data (black line), Difference pattern (blue line), Bragg positions (blue ticks).

Table 1 – Structural parameters for $Ln_3Ga_5O_{12}$, $Ln_3Sc_2Ga_3O_{12}$, $Ln_3Sc_2Al_3O_{12}$, $Ln_3In_2Ga_3O_{12}$ and $Ln_3Te_2Li_3O_{12}$; $Ln$ = Gd, Tb, Dy, Ho. All refinements were carried out in the space group $Ia\bar{3}d$, with $Ln$ on the 24$c$ sites (0, 0.25, 0.125), Ga$_{oct}$/Sc/In/Te on the 16$a$ sites (0, 0, 0), Ga$_{tetr}$/Al/Li on the 24$d$ sites (0, 0.25, 0.375) and O on the 96$h$ ($x$, $y$, $z$) sites.

| $Ln$ | | Gd[a] | Tb[a] | Dy[a] | Ho[b] |
|---|---|---|---|---|---|
| **$Ln_3Ga_5O_{12}$** | | | | | |
| $a$ (Å) | | 12.38348 (2) | 12.34191 (4) | 12.31057 (5) | 12.28157 (5) |
| $Ln$ | $B_{iso}$ (Å$^2$) | 0.5 | 0.5 | 0.5 | 0.5 |
| Ga$_{oct}$ | $B_{iso}$ (Å$^2$) | 0.5 | 0.5 | 0.5 | 0.5 |
| Ga$_{tetr}$ | $B_{iso}$ (Å$^2$) | 0.5 | 0.5 | 0.5 | 0.5 |
| O | $x$ | -0.0327 (2) | -0.0307 (3) | -0.0299 (3) | -0.02976 (2) |
| | $y$ | 0.0542 (2) | 0.0541 (3) | 0.0539 (3) | 0.05150 (3) |
| | $z$ | 0.1490 (2) | 0.1499 (4) | 0.1495 (4) | 0.1494 (3) |
| | $B_{iso}$ (Å$^2$) | 0.5 | 0.5 | 0.5 | 0.5 |
| R$_{wp}$ | | 10.2 | 9.98 | 10.1 | 9.74 |
| $\chi^2$ | | 2.86 | 2.92 | 2.59 | 4.32 |
| **$Ln_3Sc_2Ga_3O_{12}$** | | | | | |
| $a$ (Å) | | 12.57321 (7) | 12.53907 (6) | 12.50241 (6) | 12.47517 (3) |
| $Ln$ | $B_{iso}$ (Å$^2$) | 0.5 | 0.5 | 0.5 | 0.08 (2) |
| Sc | $B_{iso}$ (Å$^2$) | 0.5 | 0.5 | 0.5 | 0.19 (2) |
| Ga | $B_{iso}$ (Å$^2$) | 0.5 | 0.5 | 0.5 | 0.30 (2) |
| O | $x$ | -0.0299 (5) | -0.0310 (3) | -0.3035 (3) | -0.0283 (8) |
| | $y$ | 0.0571(5) | 0.0574 (3) | 0.0567 (3) | 0.0582 (9) |
| | $z$ | 0.1521 (5) | 0.1552 (3) | 0.1581 (3) | 0.1545 (8) |
| | $B_{iso}$ (Å$^2$) | 0.5 | 0.5 | 0.5 | 0.422 (15) |
| R$_{wp}$ | | 15.6 | 10.7 | 10.9 | 10.6 |
| $\chi^2$ | | 1.57 | 1.87 | 1.68 | 2.96 |
| **$Ln_3Sc_2Al_3O_{12}$** | | | | | |
| $a$ (Å) | | 12.43448 (6) | 12.39878 (5) | 12.35927 (6) | 12.32972 (3) |
| $Ln$ | $B_{iso}$ (Å$^2$) | 0.5 | 0.5 | 0.5 | 0.13 (2) |
| Sc | $B_{iso}$ (Å$^2$) | 0.5 | 0.5 | 0.5 | 0.17 (2) |
| Al | $B_{iso}$ (Å$^2$) | 0.5 | 0.5 | 0.5 | 0.32 (4) |
| O | $x$ | -0.0328 (3) | -0.3120 (3) | -0.0328 (3) | -0.03155 (10) |
| | $y$ | 0.5335 (4) | 0.5483 (4) | 0.5402 (3) | 0.56744 (11) |
| | $z$ | 0.1561 (4) | 0.6559 (4) | 0.1562 (4) | 0.1566 (9) |
| | $B_{iso}$ (Å$^2$) | 0.5 | 0.5 | 0.5 | 0.43 (2) |
| R$_{wp}$ | | 13.2 | 14.7 | 12.8 | 12.4 |
| $\chi^2$ | | 2.27 | 3.18 | 2.06 | 2.88 |
| **$Ln_3In_2Ga_3O_{12}$** | | | | | |
| $a$ (Å) | | 12.66112 (6) | 12.62654 (5) | 12.59268 (6) | 12.55859 (9) |
| $Ln$ | $B_{iso}$ (Å$^2$) | 0.5 | 0.5 | 0.5 | 0.11 (2) |
| In | $B_{iso}$ (Å$^2$) | 0.5 | 0.5 | 0.5 | 0.03 (3) |
| Ga | $B_{iso}$ (Å$^2$) | 0.5 | 0.5 | 0.5 | 0.37 (2) |
| O | $x$ | -0.0318 (3) | -0.0319 (3) | -0.0330 (6) | -0.0283 (7) |
| | $y$ | 0.0583 (3) | 0.0598 (3) | 0.0608 (4) | 0.0599 (8) |
| | $z$ | 0.1573 (4) | 0.1565 (4) | 0.1571 (5) | 0.1567 (8) |
| | $B_{iso}$ (Å$^2$) | 0.5 | 0.5 | 0.5 | 0.476 (15) |
| $R_{wp}$ | | 10.1 | 10.0 | 11.7 | 11.5 |
| $\chi^2$ | | 1.80 | 2.38 | 2.75 | 2.15 |
| **$Ln_3Te_2Li_3O_{12}$** | | | | | |
| $a$ (Å) | | 12.39402 (2) | 12.34898 (2) | 12.31050 (2) | 12.27401 (7) |
| Ln | $B_{iso}$ (Å$^2$) | 0.5 | 0.5 | 0.5 | 0.545 (19) |
| Te | $B_{iso}$ (Å$^2$) | 0.5 | 0.5 | 0.5 | 0.222 (24) |
| Li | $B_{iso}$ (Å$^2$) | 0.5 | 0.5 | 0.5 | 0.52 (7) |
| O | $x$ | -0.0274 (3) | -0.0255 (4) | -0.0274 (3) | -0.0264 (7) |
|  | $y$ | 0.0501 (4) | 0.0504 (4) | 0.0512 (4) | 0.0527 (8) |
|  | $z$ | 0.1450 (4) | 0.1432 (4) | 0.1457 (3) | 0.1453 (8) |
|  | $B_{iso}$ (Å$^2$) | 0.5 | 0.5 | 0.5 | 0.330 (15) |
| $R_{wp}$ | | 12.5 | 13.1 | 14.3 | 12.0 |
| $\chi^2$ | | 1.89 | 2.96 | 2.21 | 2.75 |

[a]PXRD only

[b]PXRD + PND except $Ho_3Ga_5O_{12}$

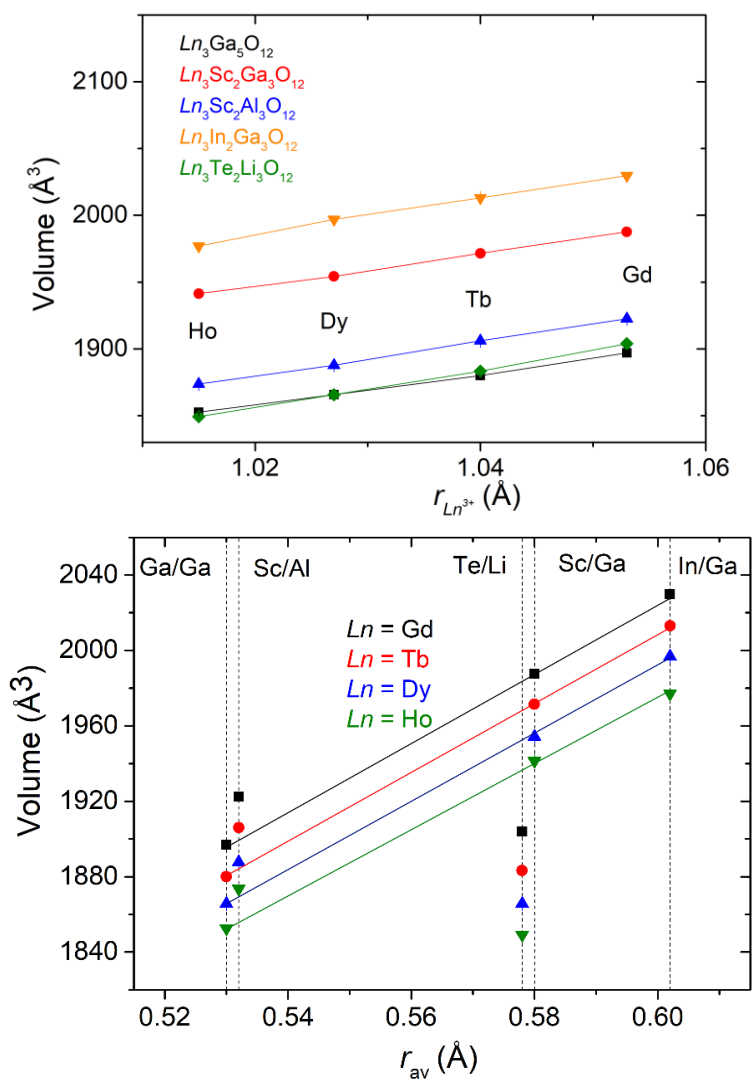

Figure 3 – Variation of lattice volume with a) ionic radii of *Ln* b) weighted ionic radii of *A* and *X*, $r_{av} = \frac{2r_A + 3r_X}{5}$

Table 2 - Selected bond lengths for $Ln_3Ga_5O_{12}$, $Ln_3Sc_2Ga_3O_{12}$, $Ln_3Sc_2Al_3O_{12}$, $Ln_3In_2Ga_3O_{12}$ and $Ln_3Te_2Li_3O_{12}$; $Ln$ = Gd, Tb, Dy, Ho.

| *Ln* | Gd[a] | Tb[a] | Dy[a] | Ho[b] |
|---|---|---|---|---|
| **$Ln_3Ga_5O_{12}$** | | | | |
| *Ln*-*Ln* (Å) | 3.79169 (5) × 4 | 3.77892 (3) × 4 | 3.76933 (3) × 4 | 3.76045 (3) × 4 |
| *Ln*-O (Å) | 2.413 (2) × 4 | 2.380 (5) × 4 | 2.368 (5) × 4 | 2.354 (4) × 4 |
|  | 2.476 (2) × 4 | 2.467 (4) × 4 | 2.461 (4) × 4 | 2.483 (3) × 4 |
| <*Ln*-O> (Å) | 2.444 | 2.424 | 2.414 | 2.418 |
| $Ga_{oct}$ –O (Å) | 2.005 (2) × 6 | 2.003 (4) × 6 | 1.991 (4) × 6 | 1.975 (3) × 6 |
| $Ga_{tetr}$-O (Å) | 1.820 (3) × 4 | 1.824 (4) × 4 | 1.828 (4) × 4 | 1.815 (4) × 4 |
| **$Ln_3Sc_2Ga_3O_{12}$** | | | | |
| *Ln*-*Ln* (Å) | 3.84974 (3) × 4 | 3.83929 (3) × 4 | 3.82807 (3) × 4 | 3.81964 (3) × 4 |
| *Ln*-O (Å) | 2.414 (7) × 4 | 2.422 (4) × 4 | 2.405 (4) × 4 | 2.372 (4) × 4 |
|  | 2.477 (7) × 4 | 2.469 (4) × 4 | 2.468 (4) × 4 | 2.495 (3) × 4 |
| <*Ln*-O> (Å) | 2.446 | 2.446 | 2.436 | 2.434 |
| Sc –O (Å) | 2.077 (7) × 6 | 2.069 (4) × 6 | 2.063 (4) × 6 | 2.074 (4) × 6 |
| Ga-O (Å) | 1.861 (7) × 4 | 1.852 (4) × 4 | 1.846 (4) × 4 | 1.815 (4) × 4 |
| **$Ln_3Sc_2Al_3O_{12}$** | | | | |
| *Ln*-*Ln* (Å) | 3.80727 (3) × 4 | 3.79634 (3) × 4 | 3.78424 (3) × 4 | 3.77464 (3) × 4 |
| *Ln*-O (Å) | 2.378 (5) × 4 | 2.361 (5) × 4 | 2.365 (5) × 4 | 2.313 (5) × 4 |
|  | 2.508 (4) × 4 | 2.481 (4) × 4 | 2.486 (4) × 4 | 2.460 (4) × 4 |
| <*Ln*-O> (Å) | 2.443 | 2.421 | 2.426 | 2.386 |
| Sc –O (Å) | 2.092 (5) × 6 | 2.085 (5) × 6 | 2.082 (4) × 6 | 2.075 (4) × 6 |
| Al-O (Å) | 1.766 (5) × 4 | 1.782 (5) × 4 | 1.757 (5) × 4 | 1.792 (5) × 4 |
| **$Ln_3In_2Ga_3O_{12}$** | | | | |
| *Ln*-*Ln* (Å) | 3.87666 (3) × 4 | 3.86608 (3) × 4 | 3.85571 (3) × 4 | 3.84282 (3) × 4 |
| *Ln*-O (Å) | 2.423 (5) × 4 | 2.427 (5) × 4 | 2.432 (6) × 4 | 2.394 (5) × 4 |
|  | 2.494 (4) × 4 | 2.467 (4) × 4 | 2.452 (6) × 4 | 2.477 (4) × 4 |
| <*Ln*-O> (Å) | 2.458 | 2.447 | 2.442 | 2.436 |
| In –O (Å) | 2.162 (4) × 6 | 2.153 (5) × 6 | 2.162 (6) × 6 | 2.154 (4) × 6 |
| Ga-O (Å) | 1.820 (5) × 4 | 1.829 (5) × 4 | 1.816 (6) × 4 | 1.796 (5) × 4 |
| **$Ln_3Te_2Li_3O_{12}$** | | | | |
| *Ln*-*Ln* (Å) | 3.79488 (3) × 4 | 3.78109 (3) × 4 | 3.76930 (3) × 4 | 3.75812 (3) × 4 |
| *Ln*-O (Å) | 2.376 (5) × 4 | 2.363 (5) × 4 | 2.360 (5) × 4 | 2.374 (4) × 4 |
|  | 2.513 (5) × 4 | 2.495 (5) × 4 | 2.484 (4) × 4 | 2.484 (3) × 4 |
| <*Ln*-O> (Å) | 2.444 | 2.429 | 2.422 | 2.429 |
| Te –O (Å) | 1.931 (5) × 6 | 1.901 (5) × 6 | 1.931 (5) × 6 | 1.892 (4) × 6 |
| Li-O (Å) | 1.882 (5) × 4 | 1.906 (5) × 4 | 1.867 (5) × 4 | 1.877 (4) × 4 |

[a]PXRD only

[b]PXRD + PND except $Ho_3Ga_5O_{12}$

## 3.2 Bulk magnetic properties

### 3.2.1 Magnetic susceptibility

The ZFC magnetic susceptibility, $\chi$, of the different lanthanide garnets as a function of temperature $T$ (2 – 300 K) measured in a field of 100 Oe is shown in Figure 4a. An ordering transition is observed for $Dy_3Sc_2Al_3O_{12}$ at $T = 2.2(1)$ K. None of the other garnets show any magnetic ordering at $T \geq 2$ K. The inverse susceptibility, Figure 4b, can be fit to the Curie-Weiss law $\frac{1}{\chi} = \frac{T - \theta_{CW}}{C}$ where $\theta_{CW}$ is the Weiss temperature and $C$ is the Curie constant. The magnetic moment, $\mu_{eff}$, is determined from the Curie constant as $\mu_{eff} = \sqrt{\frac{3 k_B C}{N_A \mu_B^2}}$. For garnets with $Ln$ = Gd, the fit to the Curie-Weiss law is carried out in the temperature range 100-300 K to calculate $\theta_{CW}$ and $\mu_{eff}$. However, for the lanthanide ions with strong single-ion anisotropy, $Ln$ = Tb, Dy, Ho, the presence of low-lying crystal electric field states[43,45,54,55] impact the values of $\theta_{CW}$, therefore the fit is carried out at low temperatures, between 2-10 K. Parameters determined from the fits are summarised in Table 3. The negative values of the Curie-Weiss temperature, $\theta_{CW}$, indicate net antiferromagnetic interactions. For $Ln$ = Gd, the experimentally determined magnetic moments derived from the Curie-Weiss fit are consistent with the theoretical free-ion values given by $\mu_{th} = g_J \sqrt{J(J+1)} \mu_B$. However, for $Ln$ = Tb, Dy, Ho they are slightly underestimated compared to the free ion values. This can be attributed to partial quenching of the angular momentum due to the crystal field; similar results have been previously reported for $Ln_3Ga_5O_{12}$, $Ln$ = Tb, Dy, Ho[38,41,42,45].

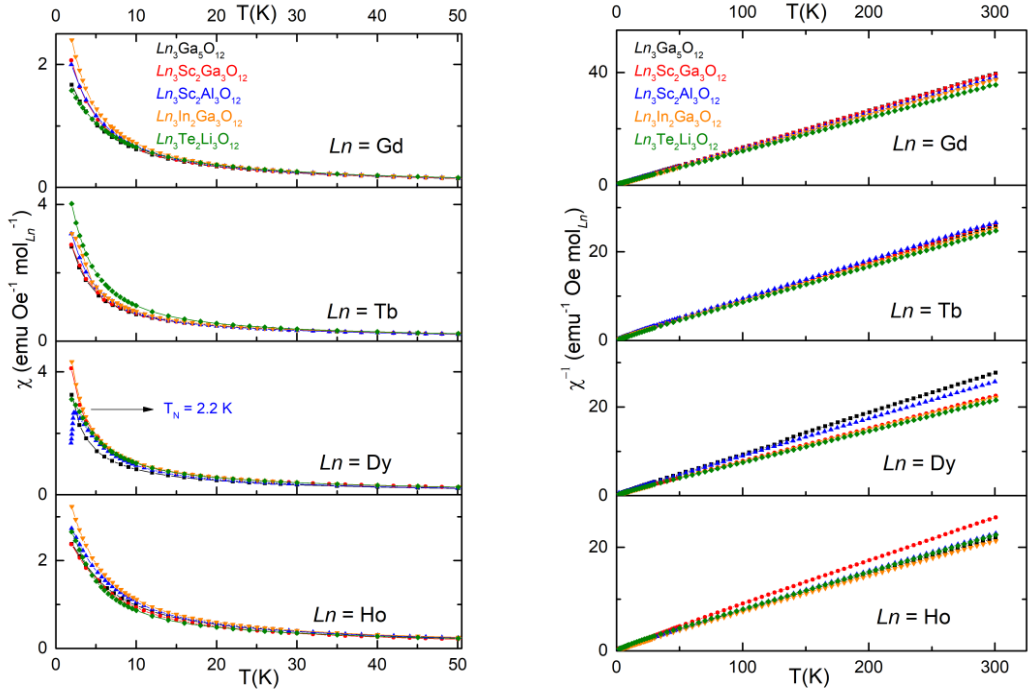

Figure 4 –a) ZFC $\chi(T)$ from 2- 300 K in a field of 100 Oe and b) inverse for $Ln_3Ga_5O_{12}$ $Ln_3Sc_2Ga_3O_{12}$, $Ln_3Sc_2Al_3O_{12}$, $Ln_3In_2Ga_3O_{12}$ and $Ln_3Te_2Li_3O_{12}$; $Ln$ = Gd, Tb, Dy, Ho

Table 3 - Parameters for Curie-Weiss fit for $Ln_3Ga_5O_{12}$, $Ln_3Sc_2Ga_3O_{12}$, $Ln_3Sc_2Al_3O_{12}$, $Ln_3In_2Ga_3O_{12}$ and $Ln_3Te_2Li_3O_{12}$ for $Ln$ = Gd, Tb, Dy, Ho; the theoretical value quoted is the free-ion value, $g_J\sqrt{J(J+1)}\,\mu_B$ where $g_J = \frac{3}{2} + \frac{S(S+1)-L(L+1)}{2J(J+1)}$

| $Ln$ | | Gd[a] | Tb[b] | Dy[b] | Ho[b] |
|---|---|---|---|---|---|
| | Theoretical $\mu_{eff}$ ($\mu_B$) | 7.94 | 9.72 | 10.65 | 10.61 |
| $Ln_3Ga_5O_{12}$ | Experimental $\mu_{eff}$ ($\mu_B$) | 7.814 (3) | 8.34 (10) | 8.34 (14) | 10.3 (3) |
| | $\theta_{CW}$ (K) | -1.4 (2) | -1.16 (4) | -1.00 (6) | -3.29 (10) |
| $Ln_3Sc_2Ga_3O_{12}$ | Experimental $\mu_{eff}$ ($\mu_B$) | 7.837 (9) | 8.50 (10) | 9.37 (10) | 9.87 (15) |
| | $\theta_{CW}$ (K) | -2.2 (3) | -1.2 (3) | -0.8 (3) | -2.9 (5) |
| $Ln_3Sc_2Al_3O_{12}$ | Experimental $\mu_{eff}$ ($\mu_B$) | 7.952 (7) | 8.5 (3) | 9.1 (3) | 10.5 (2) |
| | $\theta_{CW}$ (K) | -1.3 (3) | -0.84 (10) | -1.2 (2) | -2.91 (6) |
| $Ln_3In_2Ga_3O_{12}$ | Experimental $\mu_{eff}$ ($\mu_B$) | 8.019 (8) | 8.5 (3) | 9.48 (15) | 10.44 (13) |
| | $\theta_{CW}$ (K) | -0.6 (3) | -0.57 (11) | -0.66 (5) | -2.09 (4) |
| $Ln_3Te_2Li_3O_{12}$ | Experimental $\mu_{eff}$ ($\mu_B$) | 8.233 (4) | 9.43 (13) | 9.7 (2) | 8.98 (13) |
| | $\theta_{CW}$ (K) | -2.73 (17) | -0.63 (4) | -1.52 (7) | -1.60 (5) |

[a]Fit carried out in temperature range 100 – 300 K

[b]Fit carried out in temperature range 2 – 10 K

Table 4 – Dipolar interaction energy, $D$, and nearest-neighbour exchange interaction energy, $J_1$, for $Ln_3Ga_5O_{12}$, $Ln_3Sc_2Ga_3O_{12}$, $Ln_3Sc_2Al_3O_{12}$, $Ln_3In_2Ga_3O_{12}$ and $Ln_3Te_2Li_3O_{12}$ for $Ln$ = Gd, Tb, Dy, Ho.

| $Ln$ | | Gd | Tb | Dy | Ho |
|---|---|---|---|---|---|
| $Ln_3Ga_5O_{12}$ | $D$ (K) | 0.69 | 0.80 | 0.81 | 1.23 |
| | $J_1$ (K) | 0.51 | 0.43 | 0.38 | 1.23 |
| $Ln_3Sc_2Ga_3O_{12}$ | $D$ (K) | 0.67 | 0.80 | 0.98 | 1.09 |
| | $J_1$ (K) | 0.83 | 0.46 | 0.30 | 1.10 |
| $Ln_3Sc_2Al_3O_{12}$ | $D$ (K) | 0.71 | 0.83 | 0.96 | 1.29 |
| | $J_1$ (K) | 0.48 | 0.32 | 0.44 | 1.09 |
| $Ln_3In_2Ga_3O_{12}$ | $D$ (K) | 0.69 | 0.78 | 0.98 | 1.20 |
| | $J_1$ (K) | 0.23 | 0.21 | 0.25 | 0.78 |
| $Ln_3Te_2Li_3O_{12}$ | $D$ (K) | 0.77 | 1.02 | 1.11 | 0.95 |
| | $J_1$ (K) | 1.02 | 0.24 | 0.57 | 0.6 |

The dipolar interaction energy, $D$, can be estimated as $D = \frac{\mu_0 \mu_{eff}^2 \mu_B^2}{4\pi R_{nn}^3}$ where $R_{nn}$ is the distance between nearest-neighbour magnetic $Ln^{3+}$. The nearest-neighbour exchange interaction energy, $J_1$, can be approximated from the Curie-Weiss fit as $J_1 = J_{NN}S(S+1) = \frac{3k_B\theta_{CW}}{2n}$ where $n$ is the number of nearest neighbours surrounding one $Ln^{3+}$ ion = 4, $S$ is the effective spin quantum number and $J_{nn}$ is the scale of the interaction. For Gd, $S = 7/2$ and for $Ln$ = Tb, Dy, Ho we assumed an effective $S = ½$ state. Effective spin ½ states have been observed in $Ln_3Ga_5O_{12}$ and $Ln_3Al_5O_{12}$ for $Ln$ = Tb, Dy, Ho. However the origin of the effective $S$ = ½ state differs. For $Ln$ = Dy, it is due to the ground state being an isolated Kramer's doublet at low temperatures whereas for $Ln$ = Tb, Ho, it is an admixture of two low-lying singlet states[37,40,41,48,55]. The values of $D$ and $J_1$ are given in Table 4.

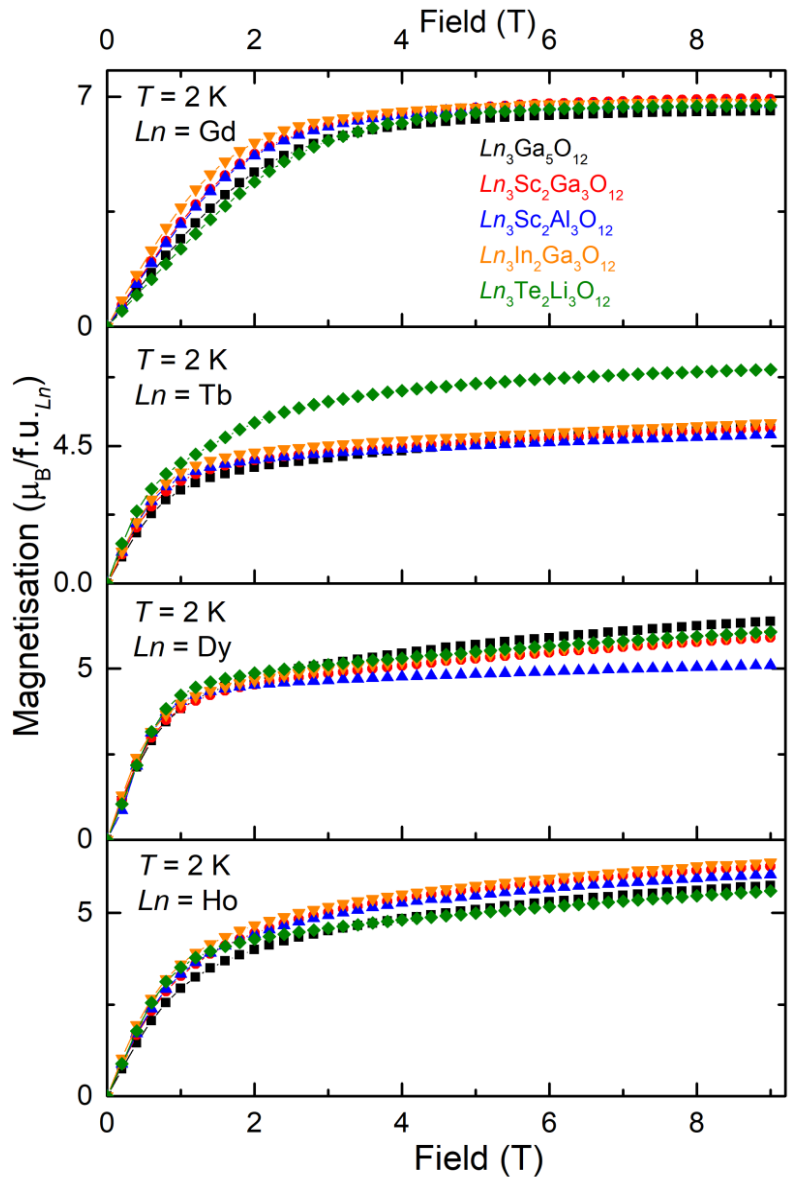

Figure 5 – $M(H)$ curves at 2 K from 0 – 9 T for $Ln_3Ga_5O_{12}$, $Ln_3Sc_2Ga_3O_{12}$, $Ln_3Sc_2Al_3O_{12}$, $Ln_3In_2Ga_3O_{12}$ and $Ln_3Te_2Li_3O_{12}$ for $Ln$ = Gd, Tb, Dy, Ho

3.2.2    Isothermal magnetisation

The isothermal magnetisation $M(H)$ at 2 K in the field range 0 – 9 T for all the garnets is shown in Figure 5 and the maximum values of magnetisation per formula unit $Ln$ (f.u._$Ln$), $M_{max}$, at 2 K, 9T are given in Table 5. The isothermal magnetisation is non-linear at low $T$ for all samples. The magnetisation of the Gd garnets saturates in a field of 9 T, with the saturation value close to 7 $\mu_B$/f.u._$Gd$ as expected for Heisenberg type spins ($M_{sat} = g_J J = 2 \times 7/2 = 7$ $\mu_B$/f.u._$Gd$). Care has to be taken when considering the isothermal magnetisation of polycrystalline samples for magnetic ions with substantial single-ion anisotropy ($Ln$ = Tb,

Dy, Ho), however these measurements have been used in other frustrated *Ln* systems to remark on the nature of the spins[56]. In the limiting field of 9 T, none of the *M(H)* curves for the garnets with *Ln* = Tb, Dy, Ho reach complete saturation. All of the observed $M_{max}$ are significantly lower than the saturation magnetisation of a Heisenberg system, $M_{sat} = g_J J$, possibly due to partial quenching of the angular momentum by the CEF, but are consistent with the magnetisation values reported for $Ln_3Ga_5O_{12}$[41,45,54,55]. We therefore postulate that the garnets with *Ln* = Tb, Dy, Ho retain strong single-ion anisotropy, however further analysis of the CEF scheme is required to determine the exact nature of the spins.

In addition, there are some subtle differences: a) the magnetisation for $Tb_3Te_2Li_3O_{12}$ is significantly higher than the other Tb garnets, possibly indicating a deviation in the single-ion anisotropy, b) among the Dy garnets, $Dy_3Sc_2Al_3O_{12}$ exhibits a greater tendency to saturate whereas a gradual increase in magnetisation with field is observed for the others. Again, this could possibly indicate a difference in single-ion anisotropy or be due to the higher Néel temperature, $T_N = 2.2(1)$ K, for $Dy_3Sc_2Al_3O_{12}$.

Table 5 - Maximum magnetisation per f.u.$_{Ln}$ for $Ln_3Ga_5O_{12}$, $Ln_3Sc_2Ga_3O_{12}$, $Ln_3Sc_2Al_3O_{12}$, $Ln_3In_2Ga_3O_{12}$ and $Ln_3Te_2Li_3O_{12}$ for *Ln* = Gd, Tb, Dy, Ho

| *Ln* | Gd | Tb | Dy | Ho |
|---|---|---|---|---|
| Theoretical $M_{sat}$ ($\mu_B$/f.u.$_{Ln}$) | 7 | 9 | 10 | 10 |
| *M* at *T* = 2 K, $\mu_0 H$ = 9 T ($\mu_B$/f.u.$_{Ln}$) | | | | |
| $Ln_3Ga_5O_{12}$ | 6.56 (2) | 5.21 (2) | 6.39 (2) | 5.74 (2) |
| $Ln_3Sc_2Ga_3O_{12}$ | 6.92 (3) | 5.08 (2) | 5.91 (2) | 6.27 (2) |
| $Ln_3Sc_2Al_3O_{12}$ | 6.77 (3) | 4.87 (2) | 5.09 (2) | 6.04 (2) |
| $Ln_3In_2Ga_3O_{12}$ | 6.82 (3) | 5.25 (2) | 6.01 (2) | 6.38 (2) |
| $Ln_3Te_2Li_3O_{12}$ | 6.73 (3) | 6.99 (3) | 6.07 (2) | 5.60 (2) |

### 3.2.3 Zero field heat capacity

Figure 6 shows the zero-field magnetic heat capacity from 0.4 – 10 K for the garnets with *Ln* = Tb, Dy, Ho (excluding $Ln_3Ga_5O_{12}$ as they are known to order below 0.4 K). Ordering temperatures, $T_0$, and estimates of the frustration index, $f = \left|\frac{\theta_{CW}}{T_0}\right|$ [34] are given in Table 6. The zero field heat capacity for $Gd_3Sc_2Ga_3O_{12}$, $Gd_3Sc_2Al_3O_{12}$ and $Gd_3In_2Ga_3O_{12}$ show no discernible ordering features down to 0.4 K (Figure S1) while the heat capacity for GGG and $Gd_3Te_2Li_3O_{12}$ has been reported elsewhere[33].

The upturn seen below 0.6-0.8 K for all the Tb garnets is due to the onset of the nuclear Schottky anomaly for $Tb^{3+}$[42]. Sharp λ type anomalies indicative of three-dimensional antiferromagnetic ordering are observed for $Tb_3Sc_2Ga_3O_{12}$, $Tb_3Sc_2Al_3O_{12}$ and $Tb_3In_2Ga_3O_{12}$. $Tb_3Te_2Li_3O_{12}$ shows a cusp at 1.04(3) K, which is more reminiscent of short-range ordering or glassy behaviour. All the Dy garnets show sharp λ type anomalies, indicative of three-dimensional antiferromagnetic ordering. The transition for $Dy_3Te_2Li_3O_{12}$ is at 1.97(5) K, consistent with previous reports of a transition around 2 K[49]. $Ho_3Sc_2Ga_3O_{12}$, $Ho_3Sc_2Al_3O_{12}$ and $Ho_3In_2Ga_3O_{12}$ show broad features in the heat capacity at T~ 2.4(2), 2.0(2) and 1.5(1) K respectively, indicative of short range magnetic correlations. It is not possible to determine if there is any transition further below 1 K due to the sharp increase from the nuclear Schottky anomaly for $Ho^{3+}$[42,57], which would mask any transition, if present. $Ho_3Te_2Li_3O_{12}$ shows a sharp λ type transition at 1.4(1) K, which points to long-range antiferromagnetic order.

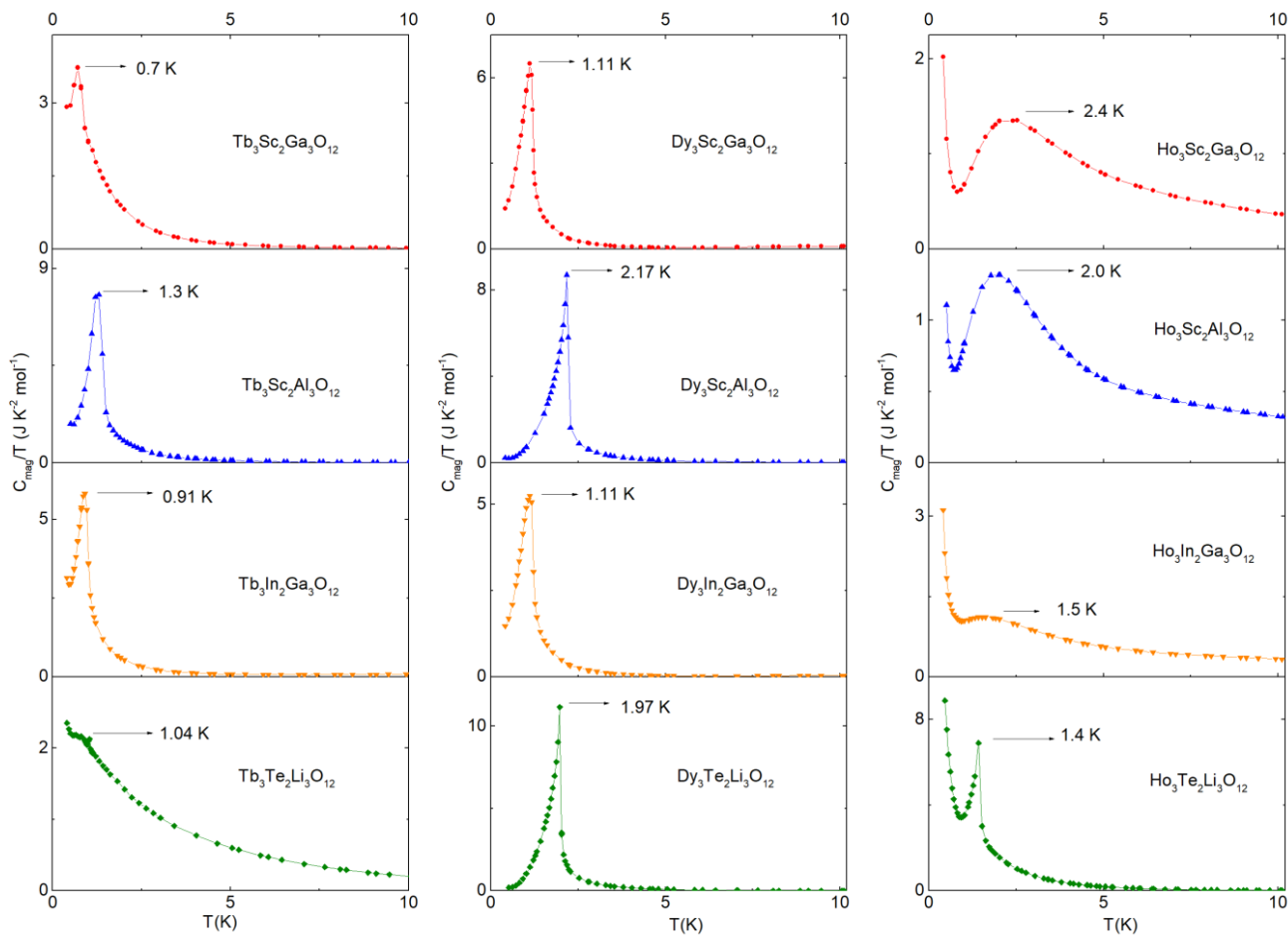

Figure 6 – Zero field heat capacity from 0.4 – 10 K for $Ln_3Sc_2Ga_3O_{12}$, $Ln_3Sc_2Al_3O_{12}$, $Ln_3In_2Ga_3O_{12}$ and $Ln_3Te_2Li_3O_{12}$, $Ln$ = Tb, Dy, Ho

Table 6 – Comparison of ordering temperature, $T_0$, and frustration index, $f$, as determined from zero field heat capacity data, for $Ln_3Sc_2Ga_3O_{12}$, $Ln_3Sc_2Al_3O_{12}$, $Ln_3In_2Ga_3O_{12}$ and $Ln_3Te_2Li_3O_{12}$ with values in the literature for $Ln_3Ga_5O_{12}$ for $Ln$ = Tb, Dy, Ho

| *Ln* | Tb | | Dy | | Ho | |
| --- | --- | --- | --- | --- | --- | --- |
| | $T_0$ (K) | $f$ | $T_0$ (K) | $f$ | $T_0$ (K) | $f$ |
| *$Ln_3Ga_5O_{12}$* | 0.25[a] | 4.6 | 0.373[b] | 2.7 | 0.19[a] | 17.3 |
| *$Ln_3Sc_2Ga_3O_{12}$* | 0.7 (1) | 1.7 | 1.11 (5) | 0.7 | 2.4 (2) | 1.4 |
| *$Ln_3Sc_2Al_3O_{12}$* | 1.3 (1) | 0.6 | 2.17 (5) | 0.5 | 2.0 (2) | 1.4 |
| *$Ln_3In_2Ga_3O_{12}$* | 0.91 (5) | 0.6 | 1.11 (5) | 0.6 | 1.5 (1) | 1.4 |
| *$Ln_3Te_2Li_3O_{12}$* | 1.04 (3) | 0.6 | 1.97 (5) | 0.8 | 1.4 (1) | 1.1 |

[a] Values taken from [42] [b] Value taken from [46]

3.2.4   Discussion

When considering the magnetic interactions, it is important to consider the impact of changes in the crystal structure. In the case of the lanthanide garnets, it is seen that the variation in $D$ for a particular $Ln$ for different combinations of $A$ and $X$ is less while $J_1$ varies more significantly (Table 4). However, these values of $D$ and $J_1$ are only an order of magnitude approximation for the respective interactions, the true values may vary considerably. The dipolar interaction is a long-ranged interaction, decaying as $1/r^3$, and so further neighbour contributions could be significant[40,58–60]. The value of $J_1$ is highly dependent on the temperature range of the fit; further neighbour exchange interactions are also likely to play a role. Despite these limitations, some qualitative conclusions may be drawn. We describe the results for each $Ln$ separately below.

$Gd_3A_2X_3O_{12}$:

$Gd_3Te_2Li_3O_{12}$ and $Gd_3Al_5O_{12}$ have been previously reported to order at 0.243 K and 0.175 K respectively while GGG shows no conventional magnetic ordering[33]. The Gd garnets are expected to show magnetic ordering, if any, at $T < 0.4$ K, below the temperature limit of our measurements. The physics of the garnets with $Ln$ = Gd should be well approximated by a microscopic Hamiltonian with $D$ and $J_1$ interaction terms; crystal field effects are expected to be less important for the isotropic $Gd^{3+}$ spins ($L=0$, $J=S=7/2$)[33]. $D$ and $J_1$ are comparable for the Gd garnets (Table 4) and the magnetic ground state would be determined mainly by the interplay of these interactions. However for GGG, it has been reported that higher order exchange interactions may play an important role[60]. Also, there may be a small single-ion anisotropy, in fact for GGG, it is the presence of both nearest neighbour antiferromagnetic correlations and a subtle XY anisotropy that is reported to lead to hidden multipolar order[32]. Thus, further neighbour contributions and subtle changes to the single-ion anisotropy caused by variation in $A$ and $X$ (especially in the case where $A$ and $X$ are aliovalent) could also impact the magnetic properties. Ultra-low temperature experimental investigations at $T < 0.4$ K as well as a more accurate modelling of the interactions is required to understand the magnetic behaviour of the Gd garnets.

$Ln_3A_2X_3O_{12}$, $Ln$ = Tb, Dy, Ho:

The dipolar and exchange interaction energies for the garnets with $Ln$ = Tb, Dy, Ho are found to be comparable. However, unlike the Gd containing garnets, $Ln$ = Tb, Dy, Ho have non-zero value of orbital angular momentum $L$ and CEF effects play an important role in addition to the dipolar and exchange interactions in determining the magnetic properties. Precise determination of the CEF requires a detailed study of the crystal field levels from inelastic neutron scattering experiments, and so only qualitative statements regarding the CEF levels and the single-ion anisotropy will be discussed here.

$Tb_3Sc_2Ga_3O_{12}$, $Tb_3Sc_2Al_3O_{12}$, $Tb_3In_2Ga_3O_{12}$ all show signatures of long-range antiferromagnetic ordering in the zero-field heat capacity with transition temperatures in between that of $Tb_3Ga_5O_{12}$ (0.25 K) and $Tb_3Al_5O_{12}$ (1.35 K)[39,41]. We propose that the ordering is still driven by interactions between the two lowest singlet states. However, the frustration is relieved as compared to $Tb_3Ga_5O_{12}$. This could be due to variations in $D$ and $J_1$ or changes in the CEF resulting in subtle changes to the single-ion anisotropy. The nature of ordering for $Tb_3Te_2Li_3O_{12}$ is very different. The transition resembles a feature observed at 0.8 K in GGG which was reported to indicate the onset of short range magnetic correlations[33]. Further experiments are required to elucidate the nature of the transition. Changes in $D$ and $J_1$ alone are not sufficient to explain the difference in the behaviour of $Tb_3Te_2Li_3O_{12}$. $Tb^{3+}$ is a non-Kramer's ion consisting of singlet levels and the energy separation between the two lowest singlet states competes with the magnetic interactions to determine the magnetic ground state. We postulate that the aliovalent $A$ and $X$ environments in $Tb_3Te_2Li_3O_{12}$ (all the other Tb garnets under consideration have trivalent $A$ and $X$) dramatically changes the single-ion anisotropy, as is also indicated from the $M(H)$ data at $T = 2$ K, causing it to behave differently to the other Tb garnets.

$Dy_3Sc_2Ga_3O_{12}$, $Dy_3Sc_2Al_3O_{12}$, $Dy_3In_2Ga_3O_{12}$ and $Dy_3Te_2Li_3O_{12}$ also show signatures of long-range antiferromagnetic ordering, with transition temperatures in between that of $Dy_3Ga_5O_{12}$ (0.373 K) and $Dy_3Al_5O_{12}$ (2.49 K)[35,46]. $Dy^{3+}$ is a Kramer's ion and so the ground state doublet is protected by symmetry for all the Dy garnets. The variation in transition temperatures and the reduction of frustration compared to $Dy_3Ga_5O_{12}$ can once again be explained by the subtle changes in the magnetic interactions and the single-ion anisotropy. $Dy_3Sc_2Al_3O_{12}$, with the maximum transition temperature, shows the greatest tendency for saturation in the $M(H)$ curves – this is consistent with the strongly Ising behaviour reported for $Dy_3Al_5O_{12}$[36].

The Ho garnets show magnetic ordering that is very different from the Tb and Dy garnets. Since the magnitude of the $D$ and $J_1$ interactions are very similar for the garnets with $Ln =$ Tb, Dy, Ho for a particular combination of $A$ and $X$, we postulate that it is the changes in CEF that causes the difference in magnetic properties. The behaviour of $Ho_3Sc_2Ga_3O_{12}$, $Ho_3Sc_2Al_3O_{12}$ and $Ho_3In_2Ga_3O_{12}$ shows similarities with $Ho_3Ga_5O_{12}$, where short range order is reported to set in below 0.6 K and long and short-range order coexist below 0.3 K[48].Therefore, it seems, the changes in CEF for $Ln =$ Ho shift the onset of short-range order to higher temperatures in the absence of long-range order at $T \geq 1$ K. Much like in the case of the Tb garnets, $Ho_3Te_2Li_3O_{12}$ behaves differently to the other Ho garnets. It exhibits a sharp λ type anomaly, indicative of three-dimensional antiferromagnetic ordering. We postulate that like $Tb_3Te_2Li_3O_{12}$, the CEF is changed dramatically for $Ho_3Te_2Li_3O_{12}$ due to the different charged environments of $Te^{6+}$ and $Li^+$.

We speculate on the possible magnetic ground states for the garnets with $Ln =$ Tb, Dy, Ho. Capel has predicted three possible kinds of magnetic ordering depending on the single-ion anisotropy (Ising, XY or Heisenberg) of the magnetic $Ln^{3+}$ ion[37]. For 4$f$ ions with a strong Ising anisotropy, type A antiferromagnetic ordering (AFA) in a six sublattice antiferromagnetic structure is predicted while type B antiferromagnetic ordering (AFB) or type C ferrimagnetic (FC) ordering is predicted for 4$f$ ions with XY or Heisenberg anisotropy[37]. This theory is directly applicable for 4$f$ ions with an odd number of electrons, like $Dy^{3+}$, where the ground state is a $S = ½$ Kramer's doublet, well-separated from excited states at sufficiently low temperatures. For non-Kramer's ions, like $Tb^{3+}$ and $Ho^{3+}$, the degeneracy is entirely lifted due to the orthorhombic point symmetry of the $Ln^{3+}$ site and thus the ground state is a singlet[37,55]. Here the energy separation between the ground state and first excited state singlet, $\Delta$, is also a factor. If $\Delta$ is much greater than the magnetic interactions, the ground state would be a non-magnetic singlet showing temperature-independent paramagnetism. However if the energy separation, $\Delta$, is small compared to the magnetic interactions, any of the three types of magnetic ordering predicted by Capel would be possible depending on the single-ion anisotropy of the $Ln^{3+}$ ion[37,38]. The sharp λ type anomalies observed in the heat capacities for the Dy garnets and majority of the Tb garnets are consistent with AFA type ordering that has also been reported for $Ln_3Ga_5O_{12}$ and $Ln_3Al_5O_{12}$ ($Ln$ = Tb, Dy, Ho) below $T_N$[36,39,43,46]. However the short range ordering observed for most of the Ho garnets and the differences in magnetic behaviour in the aliovalent $Ln_3Te_2Li_3O_{12}$ ($Ln$ = Tb, Ho) remain to be accounted for. Capel's analysis was carried out in the molecular field approximation considering predominantly dipolar interactions, further analysis of the interactions may lead to different predictions. Also, this theory does not take into account aliovalent environments (as present in $Ln_3Te_2Li_3O_{12}$) which could lead to different magnetic ground states. Neutron scattering experiments are required to elucidate the exact nature of the magnetic ground states in these garnets.

### 3.3 Magnetocaloric effect

The magnetocaloric effect (MCE) for the lanthanide garnets can be characterised by the change in magnetic entropy $\Delta S_m$ per mole $Ln$ which is calculated from the $M(H)$ curves using Maxwell's thermodynamic relation:

$$\Delta S_m = \int_{H_{initial}}^{H_{final}} \left(\frac{\partial M}{\partial T}\right)_H dH$$

Previous studies have shown that at lower fields, $\mu_0H \leq 2$ T, DGG is a better MCM whereas at higher fields, $\mu_0H > 2$ T, the change in magnetic entropy for GGG is maximised.[25] Therefore, here we compare the MCE for the garnets with substantial single-ion anisotropy ($Ln$ = Tb, Dy, Ho) in a field of 2 T and the Gd based garnets in the experimentally limiting field of 9 T. The variation in $\Delta S_m$ per mole $Ln$ as a function of temperature for all the garnets is shown in Figure 7; the inset shows the variation of $\Delta S_m$ per mole $Ln$ as a function of magnetic field. Quantitative comparisons of the MCE with the parent gallium garnets using the values of $\Delta S_m$ per mole are made in Table 7.

For $Ln$ = Gd, a 4-10% increase in $\Delta S_m$ compared to GGG (14.12 J K$^{-1}$ mol$_{Gd}^{-1}$) is observed on changing the $A$ and $X$ site ions except for Gd$_3$Te$_2$Li$_3$O$_{12}$, for which a decrease is seen. DGG has the maximum $\Delta S_m$ (3.77 J K$^{-1}$ mol$_{Dy}^{-1}$) compared to all the other Dy based garnets and so, varying the $A$ and $X$ site ions does not improve the MCE at 2 K. Among the Tb containing garnets, an increase in the MCE is observed for all compared to Tb$_3$Ga$_5$O$_{12}$ (2.88 J K$^{-1}$ mol$_{Tb}^{-1}$), with a maximum of 22.2 % increase in $\Delta S_m$ for Tb$_3$In$_2$Ga$_3$O$_{12}$ (3.52 J K$^{-1}$ mol$_{Tb}^{-1}$). However, the $\Delta S_m$ values are less than that for DGG. The most dramatic increase in $\Delta S_m$ values are observed for $Ln$ = Ho, where the MCE in Ho$_3$Te$_2$Li$_3$O$_{12}$ (3.03 J K$^{-1}$ mol$_{Ho}^{-1}$) is more than doubled compared to Ho$_3$Ga$_5$O$_{12}$ (1.41 J K$^{-1}$ mol$_{Ho}^{-1}$). Again, the absolute values of $\Delta S_m$ for the Ho garnets are less than DGG. Overall, we conclude that DGG remains the best MCM in the low field ($\leq 2$ T) regime while in a field of 9 T, the performance of GGG can be improved by changing the $A$ and $X$ site cations (~10%). If the calculations are carried out in gravimetric units that are more relevant for practical applications, a 15% and 26% increase in $\Delta S_m$ at 2 K, 9 T is observed for Gd$_3$Sc$_2$Ga$_3$O$_{12}$ and Gd$_3$Sc$_2$Al$_3$O$_{12}$ respectively as compared to GGG.

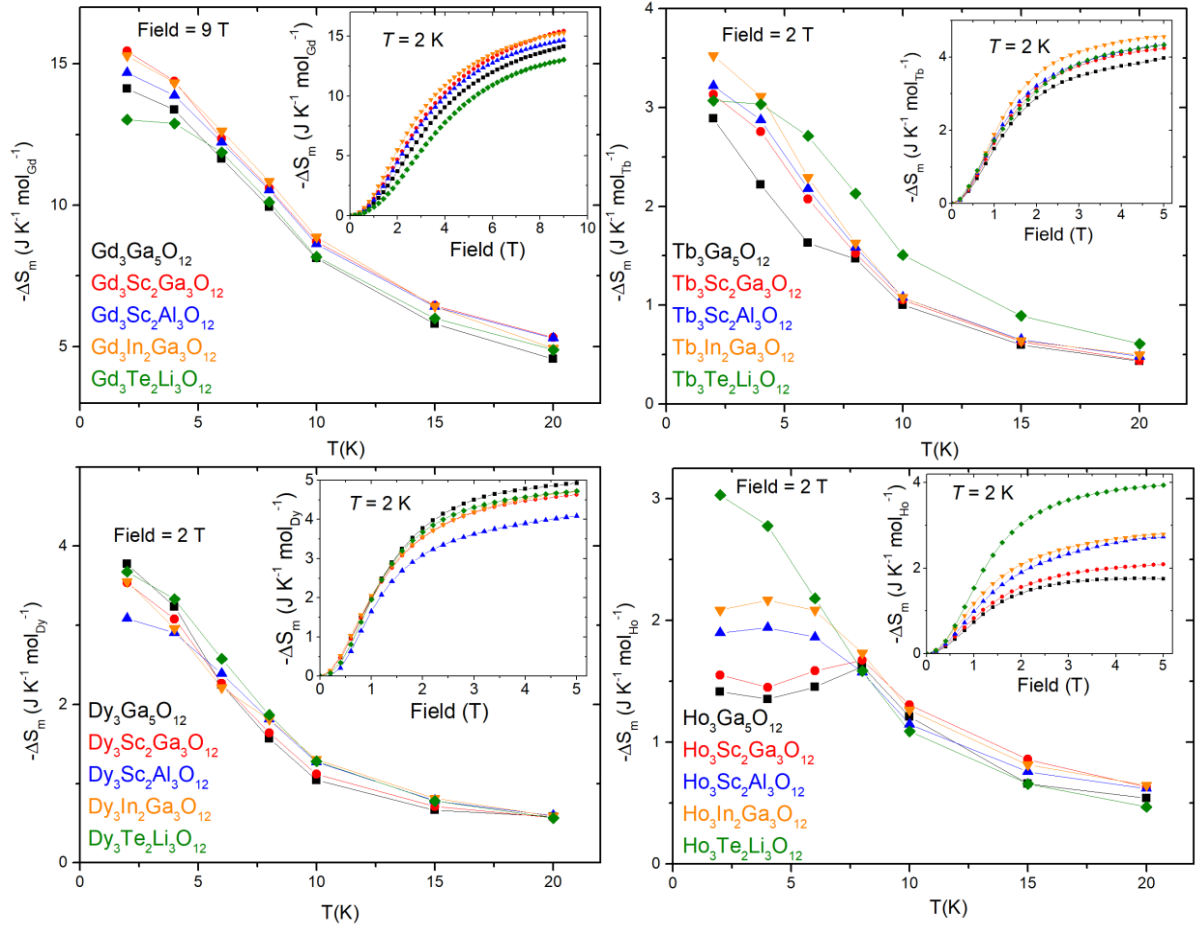

Figure 7 – $\Delta S_m$ per unit mole vs temperature; inset: $\Delta S_m$ per unit mole as a function of external magnetic field at $T = 2$ K for lanthanide garnets $Ln_3Ga_5O_{12}$, $Ln_3Sc_2Ga_3O_{12}$, $Ln_3Sc_2Al_3O_{12}$, $Ln_3In_2Ga_3O_{12}$ and $Ln_3Te_2Li_3O_{12}$, $Ln$ = Gd, Tb, Dy, Ho

Table 7 – Comparison of $\Delta S_m$ per mole $Ln$ at $T = 2$ K for $Ln_3Ga_5O_{12}$, $Ln_3Sc_2Ga_3O_{12}$, $Ln_3Sc_2Al_3O_{12}$, $Ln_3In_2Ga_3O_{12}$ and $Ln_3Te_2Li_3O_{12}$ for $Ln$ = Gd, Tb, Dy, Ho. % changes are calculated with respect to the $\Delta S_m$ values for the parent gallium garnets $Ln_3Ga_5O_{12}$

|  | Field | 9 T | | 2 T | |
|---|---|---|---|---|---|
| **Ln** | | **Gd** | **Tb** | **Dy** | **Ho** |
| **$Ln_3Ga_5O_{12}$** | $\Delta S_m$ (J K$^{-1}$ mol$_{Ln}^{-1}$) | 14.12 | 2.88 | 3.77 | 1.41 |
| **$Ln_3Sc_2Ga_3O_{12}$** | $\Delta S_m$ (J K$^{-1}$ mol$_{Ln}^{-1}$) | 15.45 | 3.13 | 3.53 | 1.55 |
| | % change | 9.4 | 8.7 | -6.4 | 9.9 |
| **$Ln_3Sc_2Al_3O_{12}$** | $\Delta S_m$ (J K$^{-1}$ mol$_{Ln}^{-1}$) | 14.68 | 3.21 | 3.09 | 1.9 |
| | % change | 4.0 | 11.4 | -18 | 35.5 |
| **$Ln_3In_2Ga_3O_{12}$** | $\Delta S_m$ (J K$^{-1}$ mol$_{Ln}^{-1}$) | 15.29 | 3.52 | 3.55 | 2.08 |
| | % change | 8.3 | 22.2 | -5.8 | 47.5 |
| **$Ln_3Te_2Li_3O_{12}$** | $\Delta S_m$ (J K$^{-1}$ mol$_{Ln}^{-1}$) | 13.02 | 3.07 | 3.68 | 3.03 |
| | % change | -7.8 | 6.6 | -2.4 | 114.9 |

## 4  Conclusion

We have prepared samples of $Ln_3Ga_5O_{12}$, $Ln_3Sc_2Ga_3O_{12}$, $Ln_3Sc_2Al_3O_{12}$, $Ln_3In_2Ga_3O_{12}$ and $Ln_3Te_2Li_3O_{12}$ for $Ln$ = Gd, Tb, Dy, Ho and measured the bulk magnetic properties. The magnetic susceptibility shows no long-range ordering down to 2 K for any of the samples except $Dy_3Sc_2Al_3O_{12}$ which orders at 2.2(1) K. Isothermal magnetisation measurements are consistent with the Heisenberg nature of the $Gd^{3+}$ spins and the substantial single-ion anisotropy reported for $Tb^{3+}$, $Dy^{3+}$, $Ho^{3+}$ in $Ln_3Ga_5O_{12}$. Garnets with $Ln$ = Tb, Dy, Ho (except $Ln_3Ga_5O_{12}$) show magnetic ordering features at $0.4 < T < 2.5$ K, however the nature of magnetic ordering varies for the different $Ln$ and combinations of $A$ and $X$. The majority of the changes in the ordering temperature can be explained by tuning of the CEF and magnetic interactions, however for non-Kramer's ions $Tb^{3+}$ and $Ho^{3+}$, in the case of $A$ and $X$ being aliovalent, other effects associated with the difference in charge must be considered.

Evaluation of the MCE shows that the magnetocaloric performance of the Gd based garnets can be improved by varying the $A$ and $X$ cations, with a ~10% increase observed for $Gd_3Sc_2Ga_3O_{12}$ compared to GGG in a field of 9 T. For the garnets with $Ln$ = Tb, Dy, Ho, DGG remains the best MCM in lower fields, $\mu_0 H \leq 2$ T.

Neutron scattering experiments to determine the magnetic ground states and CEF scheme as well as a more accurate modelling of the relevant magnetic interactions and single-ion anisotropy is required to enable a detailed comparison of the magnetic behaviour of the different garnets. We believe that this work will motivate further investigations into the magnetic properties of frustrated lanthanide garnets.

## 5  Acknowledgements

We thank Dr. Emmanuelle Suard, D2B Instrument Responsible, for her support as local contact in carrying out the neutron diffraction experiments on D2B, ILL and for valuable feedback. We thank Yutian Wu for preliminary synthetic work on $Gd_3Sc_2Ga_3O_{12}$. We acknowledge funding support from the Winton Programme for the Physics of Sustainability. Magnetic measurements were carried out using the Advanced Materials Characterisation Suite, funded by EPSRC Strategic Equipment Grant EP1M00052411.

Supporting data can be found at https://doi.org/10.17863/CAM.11904, neutron diffraction data can also be found at ILL doi:10.5291/ILL-DATA.5-31-2457.

# 7   Supporting Information

Figure S1 - Zero field heat capacity from 0.4 – 10 K for $Gd_3Sc_2Ga_3O_{12}$, $Gd_3Sc_2Al_3O_{12}$ and $Gd_3In_2Ga_3O_{12}$

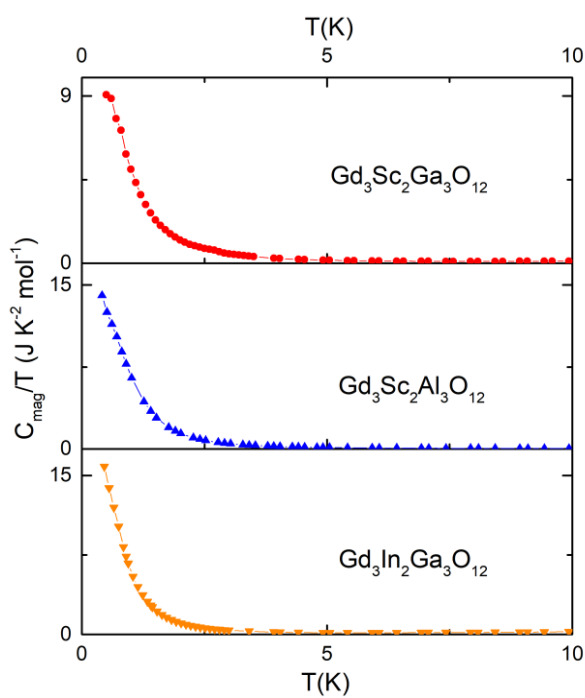